\documentclass[epj]{revtex4}
\usepackage{graphicx}
\usepackage{caption}
\usepackage{epsfig}
\usepackage{bm}
\usepackage[T1]{fontenc}
\usepackage{amssymb}
\usepackage{color}
\usepackage{float}
\usepackage{amsmath}
\usepackage[normalem]{ulem}
\usepackage{cancel}
\usepackage[colorlinks]{hyperref}
\hypersetup{
     breaklinks=true,
    pdfstartview={FitH},    % fits the width of the page to the window
    colorlinks=true,       % false: boxed links; true: colored links
    linkcolor=blue,          % color of internal links
    citecolor=red,        % color of links to bibliography
    filecolor=magenta,      % color of file links
    urlcolor=blue,           % color of external links
    anchorcolor=green,      % Color for anchor text
    linktocpage=true
}
\usepackage{microtype}
  
\begin{document}

\title{Phase transitions in four-dimensional AdS black holes with a nonlinear electrodynamics source}

\author{Ram\'on B\'ecar}
\email{rbecar@uct.cl} \affiliation{\small{Departamento de Ciencias Matem\'aticas y F\'{i}sicas, Universidad Cat\'olica de Temuco, Montt 56, Casilla 15-D, Temuco, Chile.}}
\author{P. A. Gonz\'{a}lez}
\email{pablo.gonzalez@udp.cl} \affiliation{\small{Facultad de
Ingenier\'{i}a y Ciencias, Universidad Diego Portales, Avenida Ej\'{e}rcito
Libertador 441, Casilla 298-V, Santiago, Chile.}}
\author{Joel Saavedra}
\email{joel.saavedra@ucv.cl} \affiliation{\small{Instituto de
F\'{i}sica, Pontificia Universidad Cat\'olica de Valpara\'{i}so,
Casilla 4950, Valpara\'{i}so, Chile.}}
\author{Yerko V\'{a}squez}
\email{yvasquez@userena.cl}\affiliation{\small{Departamento de F\'{\i}sica, Facultad de Ciencias, Universidad de La Serena,\\ 
Avenida Cisternas 1200, La Serena, Chile.}}
\author{Bin Wang}
\email{wang_b@sjtu.edu.cn}
\affiliation{\small{Center for Gravitation and Cosmology, College of Physical Science and Technology, Yangzhou University, Yangzhou 225009, China.}}
\date{\today}
\begin{abstract}

In this work we consider black hole solutions to Einstein theory coupled to a nonlinear power-law electromagnetic field with a fixed  exponent value. We study the extended phase space thermodynamics in canonical and grand canonical ensembles where the varying cosmological constant plays the role of an effective thermodynamic pressure. We examine thermodynamical phase transitions in such black hols and find that both first and second order phase transitions can occur in the canonical ensemble, while for the grand canonical ensemble the Hawking-Page and second order phase transitions are allowed.

\end{abstract}
\maketitle

\flushbottom

%\tableofcontents

%\newpage

%\tableofcontents

%\newpage

\section{Introduction}

Phase transitions in gravitational systems have attracted more interests since the discovery of Hawking-Page phase transitions in four-dimensional Anti-de Sitter (AdS) backgrounds, where the low-temperature phase is thermal AdS and the high temperature phase is the AdS black hole \cite{HP}.  Moreover in AdS, unlike the flat spacetime where large black holes always have smaller temperature and ordinary Schwarzschild has negative specific heat which is thermodynamically unstable, it was found that large AdS black holes have positive specific heat. If you make them bigger (higher energy), they get hotter. Small AdS black holes have negative specific heat and very small AdS black holes do not care about the cosmological constant at all so are just like the flat spacetime Schwarzschild solution. Considering  AdS black holes have the property that the horizon is a  $(d-2)$-dimensional compact Einstein space of positive, zero, or negative curvature, it was argued that they have different phase structures\cite{Birmingham:1998nr}. For Schwarzschild-AdS black holes with toroidal or hyperbolic horizons, it was found that they are thermally stable and the Hawking-Page phase transition does not occur. However, for Schwarzschild-AdS black holes with spherical horizon, there exits Hawking-Page phase transition between large stable black holes and thermal gas in the AdS space. 

Considering the cosmological constant as a thermodynamic pressure, and its conjugate variable as the thermodynamic volume, thermodynamical phase transitions have been restudied extensively from the first article about it \cite{Kastor:2009wy}, for a  review please refer to \cite{Kubiznak:2016qmn} and references therein and \cite{Belhaj:2013cva, Zeng:2016aly}. Witten \cite{Witten:1998zw} extended the four-dimensional transition to arbitrary dimension and provided a natural explanation of a confinement/deconfinement transition on the boundary field theory via the AdS/CFT correspondence.  Dynamical signature of the thermodynamical phase transition was uncovered in the study of perturbations around black holes \cite{Shen:2007xk}. In the extended phase space,  the Van der Waals like thermodynamic phase transition was further observed in quasinormal modes \cite{Liu:2014gvf}. 

Further phase transition phenomena have been analyzed and  classified by exploiting Ehrenfest's scheme \cite{Banerjee:2010da, Banerjee:2010bx,  Banerjee:2011au, Banerjee:2011raa}.   Bragg-Williams' construction of a free energy function has also been applied to examine the black hole phase transitions \cite{Banerjee:2010ve}. In addition to standard thermodynamic methods, geometrical ideas have also been introduced to study thermodynamic phase transitions and it was argued that such geometric method could disclose the intrinsic reason on the transition \cite{W,R,Ruppeiner:1995zz}.

In this work, we want to generalize the phase transition study in black hole solutions to Einstein theory coupled to a nonlinear electromagnetic field. Introducing the nonlinear electrodynamics is to eliminate the problem of infinite energy of the electron \cite{BI}. 
The nonlinear electrodynamics plays an important role 
in the construction of regular black hole solutions \cite{AyonBeato:1998ub, AyonBeato:1999rg, Cataldo:2000ns, Bronnikov:2000vy, Burinskii:2002pz, Matyjasek:2004gh}.
Some black hole/brane solutions in nonlinear electromagnetic fields have been investigated for instance in  \cite{Hendi:2010zz, Hendi:2010bk, Hendi:2010kv, Cai:2004eh, Dey:2004yt, Aiello:2004rz, Hendi:2009sw, Hendi:2013dwa, Hendi:2015hoa, Hendi:2016dmh} and references therein. The thermodynamics of Einstein-Born-Infeld black holes with a negative cosmological constant was studied in \cite{Miskovic:2008ck} and of a power-law electrodynamic hole in \cite{Gonzalez:2009nn}, where the authors  showed that a set of small black holes are locally stable by computing the heat capacity and the electrical permittivity. The thermodynamics of Gauss-Bonnet black holes for a power-law electrodynamic was studied in \cite{Hendi:2010zza}. On the other hand, higher dimensional black hole solutions to Einstein-dilaton theory coupled to Maxwell field were found in \cite{Sheykhi:2009pf, Hendi:2015xya} and black hole solutions to Einstein-dilaton theory coupled to Born-Infeld  and power-law electrodynamics were found in \cite{Dehghani:2006zi, Zangeneh:2015wia}. In our study on black hole solutions to Einstein theory coupled to a nonlinear power-law electromagnetic field, taking the specific value of the exponent power is motivated 
by the existence of exact hairy black hole solutions for the $p=3/4$ value \cite{O.:2016wcf}.   Here we want to examine phase transitions in both canonical and grand canonical ensembles.
In the study of extended phase space thermodynamics and P-V criticality of d-dimensional black holes, a nonlinear source influence on the phase structure was disclosed \cite{Hendi:2012um}. Considering the presence of a generalized Maxwell theory, such as the power Maxwell invariant, it was shown that a first order phase transition occurs in both canonical and grand canonical ensembles, in contrast to the situation in Reissner-Nordstr\"om black hole with standard Maxwell field where  phase transition only happens in the canonical ensemble.  The power Maxwell invariant (PMI) field contains richer physics than that of the Maxwell field, which reduces to linear electromagnetic source at a special case ($p=1$). The black hole solutions of the Einstein-PMI theory and their thermodynamics and geometric properties have been studied in  \cite{Hassaine:2007py,Hendi:2009zzb,Hendi:2009zza, Hassaine:2008pw, Maeda:2008ha,Hendi:2010zza,Hendi:2010bk,Hendi:2010zz,Hendi:2010kv}.  Here we will concentrate to examine how the nonlinearities, especially the power-law exponent of the nonlinear electromagnetic field, influence the phase structures of black holes.

The paper is organized as follows. In Sec. \ref{background} we give a brief review of the background that we will study. In Sec. \ref{ETD} we give an extended thermodynamics description in canonical ensembles and grand canonical ensembles. Finally, we conclude in Sec. \ref{conclusion}.

%%%%%%%%%%%%%%%%%%%%%%%%%%%%%%%%%%%%%%%%%%%%%%%%%%%%%%%%%%%%%%%%%%

\section{General Formalism for NonLinear Electrodynamics}
\label{background}
\begin{figure}[h]
\begin{center}
\includegraphics[width=0.8\textwidth]{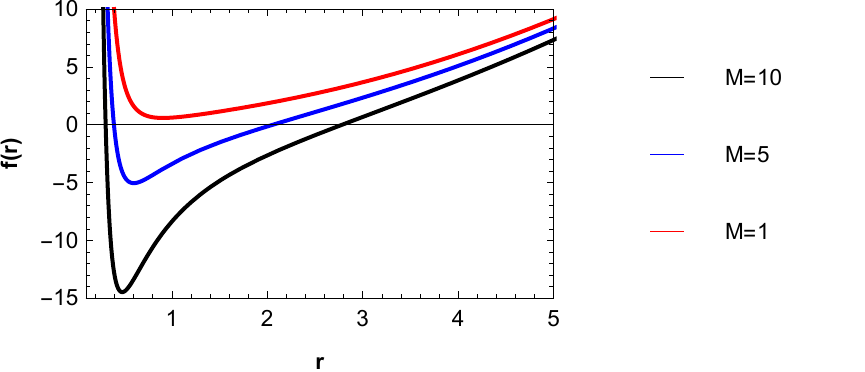}
\end{center}
\caption{The behaviour of the lapse function as a function of $r$, for different values of the black hole mass, and $q=1$, $\eta=1$ and $\Lambda=-1$}
\label{plotF}
\end{figure}

\begin{figure}[h]
\begin{center}
\includegraphics[width=0.8\textwidth]{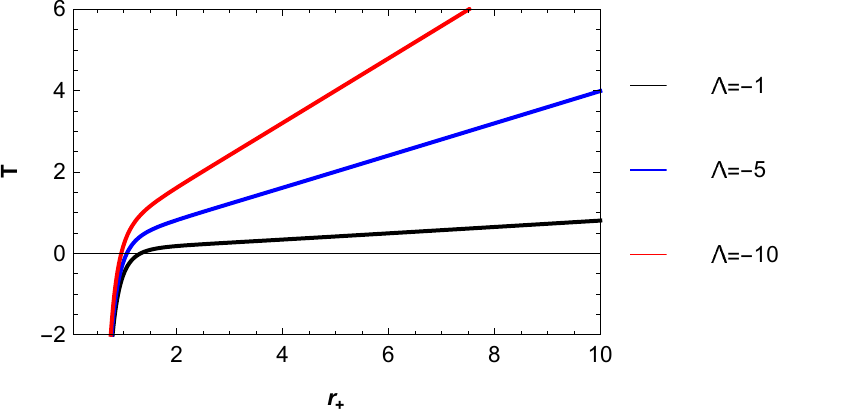}
\end{center}
\caption{This figure shows Temperature  for $k=1$, $Q=2$ and  $\eta=1$, for different values of the cosmological constant.}
\label{plotT}
\end{figure}

 We consider  topological black hole solutions for the power Maxwell theory coupled to gravity described by the action, see Refs. \cite{Hendi:2012um, Zangeneh:2015wia} 
 \begin{eqnarray} \label{action2}
 I = \int d^{4}x\sqrt{-g} \left ( \frac{1}{2 \kappa }  (R-2 \Lambda )+\eta \, |-F_{\mu \nu }F^{\mu \nu}|^p \right)\,,
 \end{eqnarray}
 where $\kappa=8 \pi G=1$, $F_{\mu \nu }=\partial_{\mu} A_{\nu}-\partial_{\nu} A_{\mu}$ and $A_{\mu}$ represents the gauge potential. The exponent $p$ is a rational number and the absolute value ensures that any configuration of electric and magnetic fields can be described by these Lagrangians. One could also consider the Lagrangian without the absolute value and the exponent $p$ restricted to being an integer or a rational number with an odd denominator \cite{Hassaine:2008pw}. The sign of the coupling constant $\eta$ will be chosen such that the energy density of the electromagnetic field is positive. This condition is guaranteed in the following cases: $p > 1/2$ and $\eta > 0$ or $p < 1/2$ and $\eta < 0$ \cite{O.:2016wcf}. Here, we will consider the first condition and a specific value of the exponent $p=3/4$. In Ref. \cite{O.:2016wcf}  was shown that the field equations have as solutions the topological nonlinearly charged black holes  Refs. \cite{O.:2016wcf}, \cite{Hendi:2012um}, \cite{Zangeneh:2015wia}.
The general AdS solution for power Maxwell black hole 
can be written as follow 
\begin{equation}
   ds^2=-f(r)dt^2+\frac{dr^2}{f(r)}+r^2d\Omega^2 \,,
\end{equation}
where $d\Omega^2$ is the metric of the spatial 2-section, which can have a positive ($k=1$), negative ($k=-1$) or zero ($k=0$) curvature, and
\begin{equation}
   f(r)= k-\frac{m}{r}- \frac{\Lambda}{3}  r^2+\eta \frac{ 2^{p} (2 p-1)^2}{(3-2p)}\frac{q^{2p}}{r^{\frac{2}{2p-1}}} \,,    \label{NLPM}
\end{equation}
where $m$ and $q$ are integration constants related to the ADM mass $M$ and the electric charge $Q$ of the black hole by 
\begin{eqnarray}
M&=&4\pi m\,, \\
Q&=& \eta  2^{2 p-1} p \, q^{2 p-1}\,,
\end{eqnarray}
respectively, and the gauge potential is given by
\begin{equation}
   A_t(r) = \frac{1-2p}{3-2p}\, q \,  r^{\frac{3-2p}{1-2p}} \,.
\end{equation}
In Ref. \cite{O.:2016wcf} it has been shown that Eq. (\ref{NLPM}) describes a black hole solution with an inner horizon ($r_{-}$) and an  outer  horizon ($r_+$). The outer horizon $r_+$ of this black hole can be calculated numerically by finding the largest real positive root of $f(r = r_+) = 0$, for different values of $p$ parameter under consideration. 
In the  following, we focus our attention at the specific value  $p=3/4$.
Then, the metric (\ref{NLPM}) reads as follows 
\begin{equation}
    f(r)=k-\frac{m}{r}+\eta\frac{  q^{3/2}}{3 \sqrt[4]{2} r^4}-\frac{\Lambda  r^2}{3}\,. \label{p34}
\end{equation}
In Fig. \ref{plotF} we show the behaviour of this black hole metric function for different values of the cosmological constant and $k=1$ with $p=\frac{3}{4}$.  We can observe that  for  a  fixed values of the cosmological constant, black hole charge, and the non-linear coupling constant $\eta$, there are different kind of solutions; black holes with a Cauchy and an event horizon, an extremal black hole configuration, and also naked  singularities,  when the black hole mass decrease.
Now, using the surface gravity relation, and Eq. (\ref{p34}) we can obtain the Hawking temperature of the black hole solutions,  and also we can express thermodynamics quantities as function of the event horizon and physical charge rather than physical mass $M$ 
\begin{equation}\label{TNLM}
 T= \frac{1}{4\pi r_+}\left(k-\eta \frac{  q^{3/2}}{ \sqrt[4]{2}  r_+^4}-\Lambda  r_+^2 \right)\,. \end{equation}
The behaviour of the temperature of this black hole can be seen in Fig. \ref{plotT}. Note that the temperature vanishes for  $r_+=r_{extrem}$, and for $r_+<r_{extrem}$ we obtain a negative temperature and therefore it corresponds to a region with nonphysical meaning where the thermodynamics description breakdown.
 Besides, the electric potential $\Phi$, measured at infinity with respect to the horizon is 
\begin{equation}
\Phi=-(A_t(r_+)-A_t(\infty))=\frac{q}{3 r_+^3}\,,
\end{equation}
while the Bekenstein-Hawking entropy $S$ is given by
\begin{equation}
S=8 \pi^{2} r_{+}^2\,.
\end{equation}
The next section is devoted to study the extended thermodynamics properties and its particular P-V criticality.

\section{Extended Thermodynamics Description}
\label{ETD}
Now at the extended  thermodynamics space we consider the cosmological constant as another thermodynamics variable, and therefore  the standard extensive parameters will be the entropy, the black hole charge, and the cosmological constant. In order to see this extended thermodynamics we consider the cosmological constant as source of a dynamical pressure using the relation 
$P=-\frac{\Lambda}{8 \pi}$ \cite{Dolan:2010ha, Dolan:2011xt}, and the conjugate thermodynamics volume is given by $V=\frac{32 \pi^{2}}{3}r_{+}^3$. After obtaining these quantities we verify that they satisfy the following Smarr formula
\begin{equation}
    M=2 T S+\frac{4}{3}\Phi Q-2 VP\,,
\end{equation}
with $M$ given by
\begin{equation} \label{MNLM}
M= 4 \pi r_+{} \left( k + \eta \frac{ 2^{3/4}     q^{3/2}}{6 r_{+}^4}-\frac{1}{3}   \Lambda  r_{+}^2 \right).
\end{equation}

\subsection{Canonical Ensemble (Fixed $Q$)}
\begin{figure}[h]
\begin{center}
\includegraphics[width=0.7\textwidth]{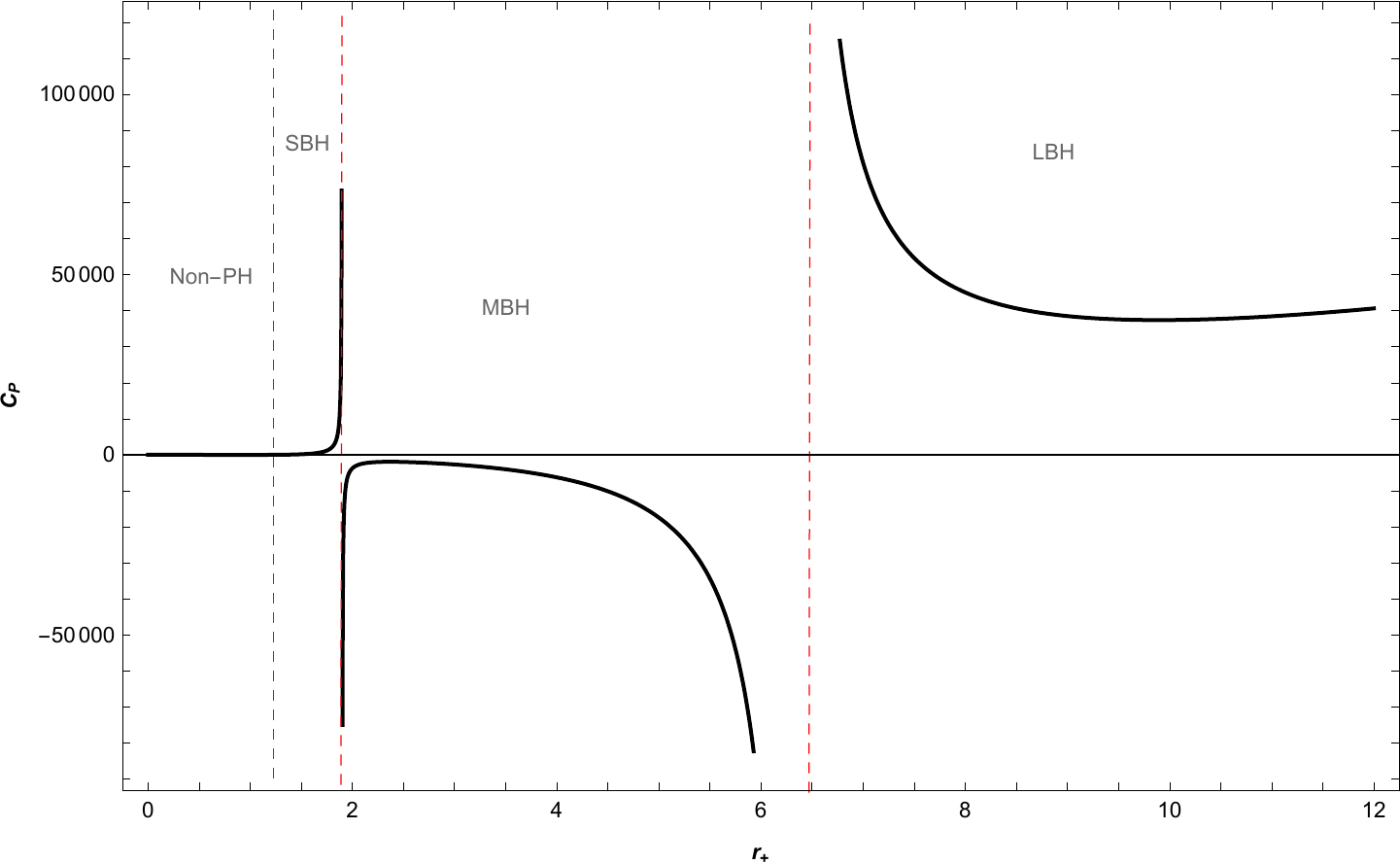}
\end{center}
\caption{The heat capacity  for $k=1$, $Q=1$, $\eta=1$ and $\Lambda=-0.0246$. This figure depict different regions according to the sign of the heat capacity; I) $C_Q<0$ and $T<0$ corresponds to non-physical region (Non-PH) , II) $T>0$: $C_Q>0$ small black hole region (SBH), $C_Q<0$ meta-stable  black hole region (MBH) and $C_Q>0$ large black hole region (LBH). The second order phase transitions occurs at the points where the heat capacity shows a divergence.}
\label{plots0}
\end{figure}

In this section we study phase transitions at the canonical ensemble, where we are considering the fixed charge as an extensive parameter. First, we compute the heat capacity for a fixed $Q$
\begin{equation}
    C_Q=T \left(\frac{ \partial S} {\partial T}\right)_Q\,.
\end{equation}
So, from Eq. (\ref{TNLM}) the
heat capacity yields
\begin{equation}\label{CNLM}
    C_Q=-\frac{16 \pi ^2 r_+^2 \left(-2 k r_+^4+2^{3/4} \eta  q^{3/2}+2 \Lambda  r_+^6\right)}{5\ 2^{3/4} \eta  q^{3/2}-2 r_+^4 \left(k+\Lambda  r_+^2\right)}\,. 
\end{equation}
Now, from Eqs. (\ref{TNLM}) and (\ref{CNLM}) we can see that the temperature and the heat capacity are always positive with the only exception for the extreme configuration where $r_+=r_-= r_{extreme}$. There is a region where $C_Q < 0$ and according to Davies’s approach this could indicated the presence of a type one phase transition, also it is important to remark that the temperature in this region is also negative and therefore it is considered as a non physical region, where the thermodynamics description breaks down.
Therefore, the main conclusion in according to Davies’s approach it is  the correlation between drastic change for the stability properties of a thermodynamic black hole system and the change of the sign or divergences of the heat capacity. In reference to the canonical ensemble, it is very well known that black holes are locally stable thermodynamics systems if its heat capacity is positive or non vanishing. Besides, at the points where the heat capacity is  divergent there is a type two phase transition. From Fig. \ref{plots0} we can see that this black hole solution is locally unstable from a thermal point of view, because the heat capacity $C_Q$ has a  divergent term, for the case where the spatial section is spherical ($k=1$). For the hyperbolic ($k=-1$) and flat ($k=0$) spatial sections we can conclude that the respective configurations of black hole are locally stable.
Then, following Davies and according to Ehrenfest’s classification, second order phase transitions occur at those points where the heat capacity diverges. These points can be obtained  from the following relation
\begin{equation}\label{condiitionCQ}
5\,* 2^{3/4} \eta  q^{3/2}-2 r_+^4 \left(k+\Lambda  r_+^2\right)=0. \,.
\end{equation}
Here, we can identify three regions separated by two divergent points. Then, we can recognize  three different phases, two stables phases where $C_Q>0$ for small and large horizon radius, and one unstable phase $C_Q<0$ for intermediate radius. Therefore, there are
three different phases that we call small black hole (SBH), meta-stable black hole (MBH) and large black hole (LBH), some results have been reported for charged AdS black holes with Maxwell and power Maxwell invariants in  \cite{Majhi:2012fz,Poshteh:2016rwc,Nam:2018sii}. 
Now, in order to figure out the second order phase transition and its consequences we study the critical behaviour performing the comparison of the equations of state with the Van der Waals equation. First using Eqs. (\ref{TNLM}),  (\ref{MNLM}) and the definition of pressure due to the cosmological constant  for a fixed charge $Q$, we obtain the equation of state, $P(V,T)$
\begin{equation}. \label{sta}
P=\frac{T}{2 r_+}-\frac{k}{8 \pi  r_+^2}+\eta \frac{  q^{3/2}}{8 \sqrt[4]{2} \pi  r_+^6}\,,
\end{equation}. 
where $r_+$ is the horizon radius. 
Following \cite{Gunasekaran:2012dq}, we identify
the geometric quantities $P$ and $T$ with physical pressure and temperature of the system by using dimensional analysis $l_{P}^2=G_{d}\hbar/c^3$. Then, we can identify the following relations between the geometric quantities $P$ and $T$ and the physical pressure and temperature:
\begin{equation}
    [Press]=\frac{\hbar c}{l_{P}^2}[P]\,, \,\,\,\,\, [Temp]=\frac{\hbar c}{k}[T]\,.
\end{equation}
Then,
\begin{equation}
    [Press]=\frac{\hbar c}{l_{P}^2}[P]=\frac{\hbar c}{l_{P}^2}\frac{ T}{ 2 r_+}+ \dots  \,,
\end{equation}
from this expression we can identify the specific volume $v$ of the fluid with the horizon radius of the black hole as $v=2 r_+$, thus, Eq. (\ref{sta}) can be written as
\begin{equation}
P=\frac{T}{v}-\frac{k}{2 \pi  v^2}+4 \eta \frac{ 2^{3/4}   q^{3/2}}{\pi  v^6}\,.
\end{equation}

\begin{figure}[h]
\begin{center}
\includegraphics[width=0.7\textwidth]{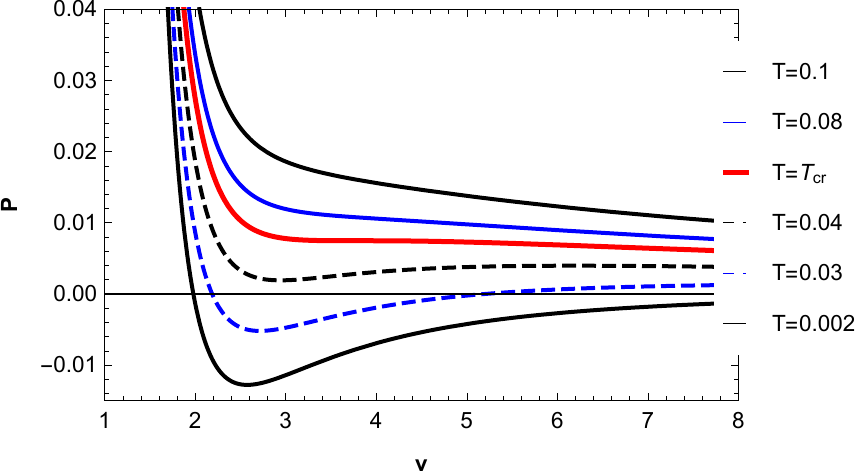}
\end{center}
\caption{ $P$-$v$ diagram with $k=1$, $q=1$ and  $\eta=1$, for different values of the temperature.}
\label{plotsPV1}
\end{figure}

In Fig. {\ref{plotsPV1}} we show the $P$-$v$ diagram, and it is possible to observe a  similar behaviour than the  Van der Waals gas,
that means, there is a critical point corresponding to an inflection point over the critical isotherm curve. 
The critical point can be computed through 
\begin{eqnarray}
\notag \frac{\partial P}{\partial v}=0\,, \\
\frac{\partial^2 P}{\partial v^2}=0\,,
\end{eqnarray}
which yields
\begin{eqnarray}\label{pointcrit}
\notag  v_{critical}&=&\frac{2^{15/16} \sqrt[4]{15} \sqrt[4]{\eta } q^{3/8}}{\sqrt[4]{k}}\,, \\
\notag T_{critical}&=&\frac{2 \sqrt[16]{2} k^{5/4}}{5 \sqrt[4]{15} \pi  \sqrt[4]{\eta } q^{3/8}}\,,\\
P_{critical}&=&\frac{k^{3/2}}{6\ 2^{7/8} \sqrt{15} \pi  \sqrt{\eta } q^{3/4}}\,.
\end{eqnarray}
This critical point only exists for black hole with spherical spatial section ($k=1$), also we can compute the universal relation 
\begin{equation}
    \rho_0=\frac{P_{critical} \times v_{critical}}{T_{critical}}=\frac{5}{12}\,.
\end{equation}
This universal number corresponds to the critical number and it does not depend on the black hole charge or the non-linear coupling constant $\eta$. This universal number differs from the Van der Waals fluid, however the critical point belongs to the same universality class as the one of the Van der Waals fluid. Now we would like to discuss the P-V criticality and the phase transition of charged black holes considering the power Maxwell electrodynamics with $p=\frac{3}{4}$ in the canonical (fixed $Q$) ensemble. In this ensemble it is possible to compute the Gibbs free energy evaluating the Euclidean action on-shell. For doing this, we have to use the counterterm method in order to cancel the divergences appearing in the on-shell Euclidean action,
\begin{equation}
I_E=I_{bulk}+I_{ct}+I_{GH}+I_{A}\,,    
\end{equation}
where $I_{bulk}$ corresponds to the Euclidean on-shell bulk action, $I_{ct}$ is the action coming from the counterterm method, $I_{GH}$ is the well known Gibbons-Hawking term and $I_{A}$ is a boundary term for the electromagnetic field needed for a well posed action principle, this term in the canonical ensemble (fixed $Q$) is not vanishing, and it
 is given by
 \begin{equation}
     I_{em}=-\frac{p}{4 \pi}\int _{\partial M}d^n x \sqrt{h} (-F)^{s-1} n_{\mu}F^{\mu \nu}A_{\nu}=\beta Q \Phi\,,
 \end{equation}
 where $\beta=\frac{1}{T}$ and $h_{ij}$ is the induced metric at the boundary. So, the Gibbs free energy or thermodynamics potential is given by
 \begin{equation}
     G(T,Q,P)=\frac{I_{E}^{on-shell}}{\beta}=M-TS\,,
 \end{equation}
 and the corresponding expression in our case is given by
\begin{equation}
G=2 \pi  r_+ \left(k-\frac{8}{3} \pi  P r_+^3+\eta \frac{5   q^{3/2}}{3 \sqrt[4]{2} r_+^3}\right)\,.
\end{equation}

\begin{figure}[h]
\begin{center}
\includegraphics[width=0.6\textwidth]{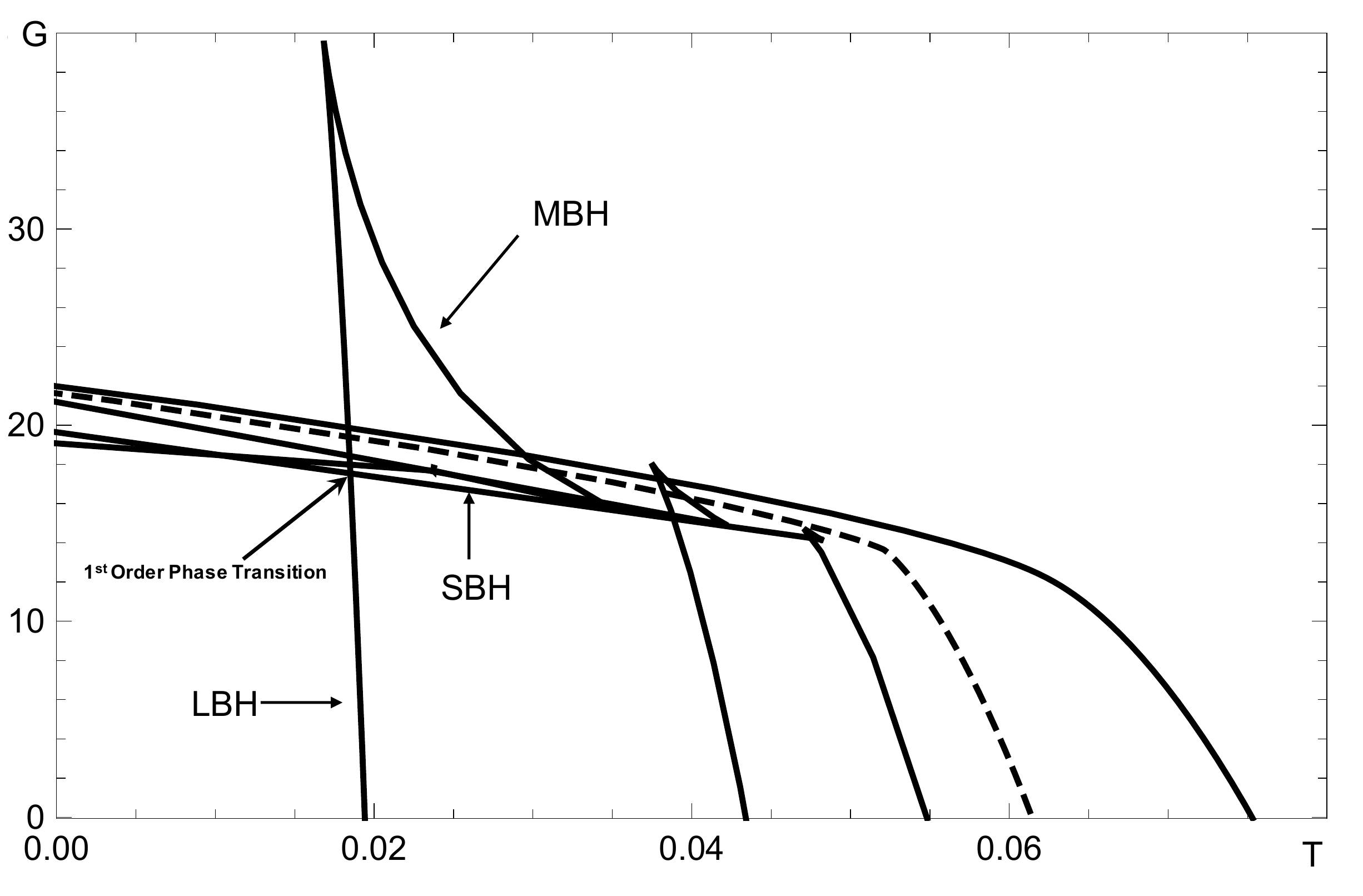}
\end{center}
\caption{Gibbs free energy for nonlinear Maxwell electrodynamics for $p=\frac{3}{4}$, $k=1$, $q=2$ and $\eta=1$ at the canonical ensemble. We can see that for $P<P_C$ ($P_C$ is represented by the dashed line) the system is undergoing by a first order phase transition described by the characteristic swallowtail}
\label{plots00}
\end{figure}
In Fig. \ref{plots00}, we depict of Gibbs free energy, clearly the critical pressure is marking the point where the system is undergoing by a first order phase transition. For $P<P_C$ it appears the well known characteristic swallowtail where the first order phase transition  is occurring at the intersection point between the small and large black holes configurations. For those branches we can verify that $T>0$ and  $C_Q>0$, it does mean that SBH and LBH are locally stable or also can be interpreted as the phases with positive specific heat in the lower radius and higher radius regions are stable. Now, for $T>0$ and  $C_Q<0$ we have the MBH configuration and this means the MBH branch is unstable. This meta-stable configuration can be explained because  states of the lowest Gibbs free energy are preferred by the system, similar situations can occur for some states on the small and large black hole branches.
From Eq. (\ref{condiitionCQ}) we can obtain the value of the point where the heat capacity diverges, which is exactly the critical point (\ref{pointcrit}). The first order phase transition takes place between the stable branches, and when $P=P_C$ the two divergent points merge to form a single divergence, removing the unstable region and where the second order phase transition, described by $C_Q$, occurs. For $P>P_C$ the heat capacity is positive definite and free of divergences, then the black hole configurations are stable and free of critical behaviour.

Note that due to the conservation of electric charge there is no Hawking-Page (HP) phase transition, for the charged black hole with a Gibbs free energy essentially non vanishing to the thermal AdS space with a vanishing Gibbs free energy. In the next section, we consider the grand canonical ensemble where the charge $Q$ is not fixed and therefore we will be able to study the HP phase transition. 

\subsection{Grand Canonical Ensemble (Fixed $\Phi$)}
\begin{figure}[h]
\begin{center}
\includegraphics[width=0.8\textwidth]{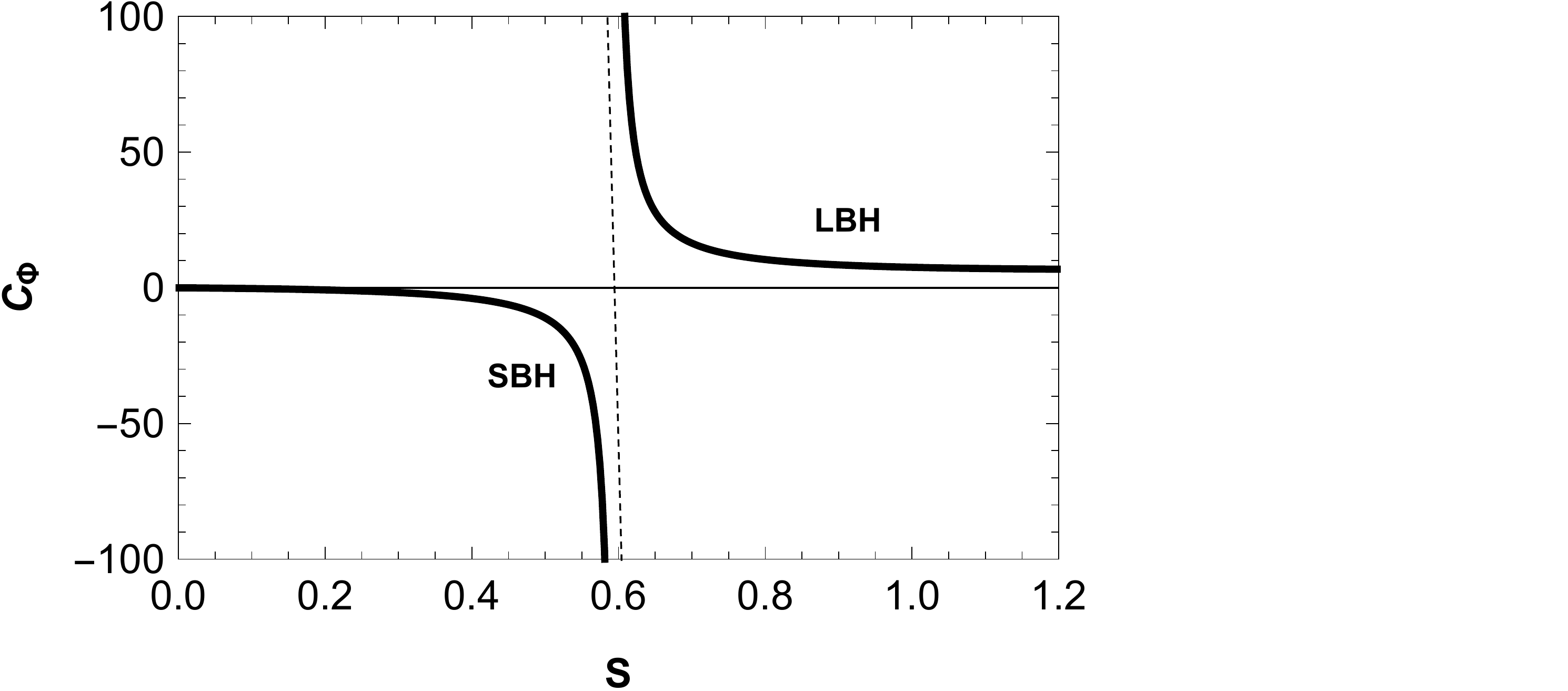}
\end{center}
\caption{Heat capacity as a function of the entropy for $\Phi=0.5$, $k=1$,  $\eta=0.1$, and $P=5$.}
\label{plots2}
\end{figure}

\begin{figure}[h]
\begin{center}
\includegraphics[width=0.6\textwidth]{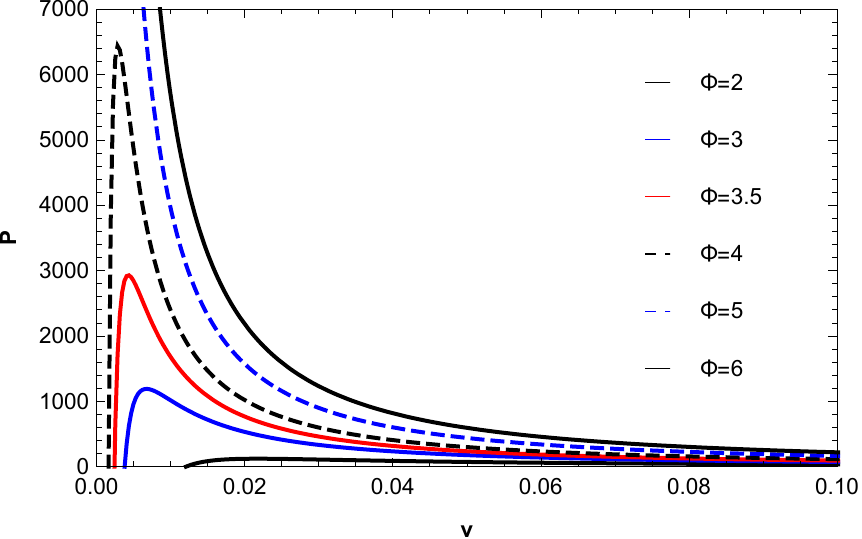}
\end{center}
\caption{$P$-$v$ diagram for $k=1$, $q=1$ and  $\eta=1$, for different values of the electric potential.}
\label{plots000}
\end{figure}

Now, in this section we discuss the P-V criticality  of charged black holes considering  the power Maxwell electrodynmaics with $p=\frac{3}{4}$ in the grand canonical (fixed $\Phi$) ensemble. 
As in the previous section, an important quantity is the heat capacity $C_{\Phi}$ at fixed $\Phi$, which is relevant to characterize the local stability in the grand canonical ensemble, and it is given by
 \begin{equation}
 C_{\Phi}=\frac{4 S\left(2 PS+2 k\pi-3 \times 2^{\frac{1}{4}}\sqrt{3\pi}S^{\frac{1}{4}}\eta\Phi^{\frac{1}{2}}\right)}{4PS-4k \pi+3 \times 2^{\frac{1}{4}}\sqrt{3\pi}S^{\frac{1}{4}}\eta\Phi^{\frac{1}{2}}}\,.\end{equation}
In Fig. \ref{plots2}, we plot the behaviour of the heat capacity as a function of the entropy, and we can conclude that there is a second order phase transition at the point where the heat capacity  diverges. In fact, at the divergence point the system is undergoing by a second order phase transition between unstable small black hole configurations ($C_{\Phi}>0$) and stable large black hole configurations ($C_{\Phi}>0$), as we will see in the following.
In this ensemble it is possible to compute the grand canonical potential evaluating the Euclidean action on-shell. In order to regularize the divergences that appear in the on-shell Euclidean action we have to use the counterterm method. However in the grand canonical ensemble $I_{A}$ is vanishing, because the electric potential $\Phi$ is kept fixed, and the grand potential becomes
\begin{equation}
    \mathcal{G}=M-TS-Q \Phi\,.
\end{equation}
Now, using the specific volume $v=2r_+$, we can obtain the temperature in this ensemble, which reads as follows
\begin{equation}
    T=\frac{1}{4 \pi  r_+}\left( k+8 \pi  P r_+^2-\eta \frac{3 \sqrt{3} (r_+ \Phi ^3)^{\frac{1}{2}}}{\sqrt[4]{2}} \right)\,,
\end{equation}
 and the equation of state is given by
 \begin{equation} \label{PGC}
     P=\frac{T}{v}-\frac{k}{2 \pi  v^2}+\eta \frac{3 \sqrt{3} \Phi ^{3/2}}{2\ 2^{3/4} \pi  v^{3/2}}\,.
 \end{equation}
Then, from Eq. (\ref{PGC}) and the conditions $\partial P=\partial^2 P=0$, the critical point yields
\begin{eqnarray}\label{pointcr}
\notag v_{critical}&=&\frac{128\sqrt{2} k^2}{243 \eta^2\Phi^3}\,, \\
\notag T_{critical}&=&-\frac{243\eta^2\Phi^3}{128\sqrt{2}k\pi}\,,\\
P_{critical}&=&\frac{19683 \eta ^4 \Phi ^6 \left(8 \eta  \Phi ^{3/2} \sqrt{\frac{k^2}{\eta ^2 \Phi ^3}}-9 k\right)}{65536 \pi  k^4}\,.
\end{eqnarray}
Note that, the critical point has physical meaning only for black hole with hyperbolic spatial section ($k=-1$) and lose physical meaning for $k=1$ and $k=0$, this point was argued in Ref. \cite{Hendi:2012um} for $p<1$. Then, for these cases, the P-v diagram does not show a critical point because the condition $\partial P=\partial^2 P=0$ can not be satisfied and therefore there is not a critical point in the gran canonical ensemble for $k=1$ and $k=0$. Therefore, it is more like a solid–liquid phase transition, rather than a liquid–gas phase transition. However, for an interval of the electric potential the pressure has a maximum $P_{max}$, these results are summarized in Fig. \ref{plots000}. Also, for $k=-1$ we can compute the universal relation $\rho_0=\frac{P_{critical}*v_{critical}}{T_{critical}}$  for the gran canonical ensemble, giving $ \rho_0= 1/6$.

\begin{figure}[h]
\begin{center}
\includegraphics[width=0.6\textwidth]{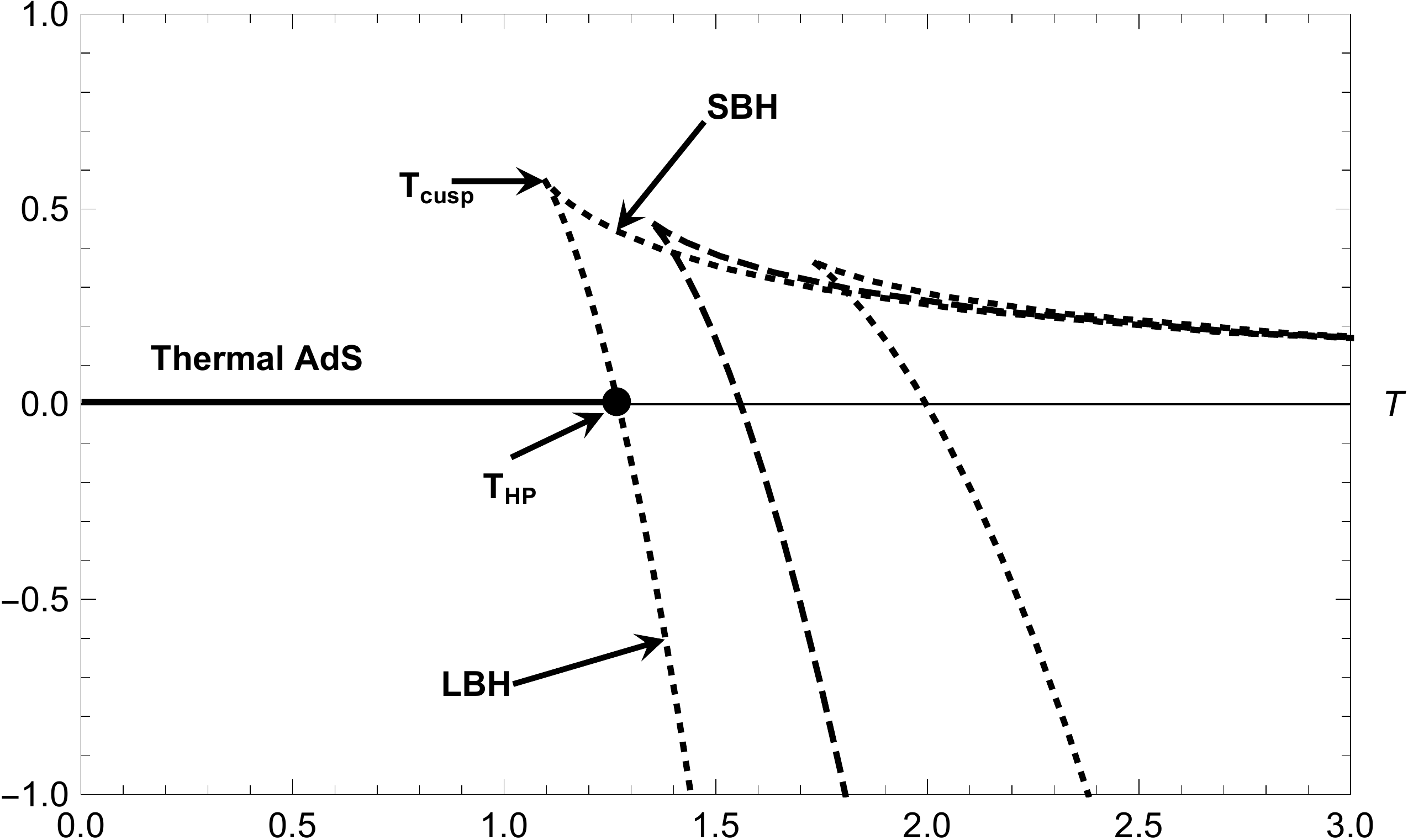}
\end{center}
\caption{Gibbs free energy for nonlinear Maxwell electrodynamics for $p=\frac{3}{4}$, $k=1$,  $\eta=0.1$ for different values of the electric potential $\Phi$ at the grand canonical ensemble}
\label{plotsG}
\end{figure}

On the other hand, the thermodynamic grand potential can be written as
 \begin{equation}
    \mathcal{G}= -\frac{1}{3} \pi  \left(-6 k r_++16 \pi  P r_+^3+3\ 2^{3/4} \sqrt{3} \eta  \Phi  \sqrt{r_+^3 \Phi }\right).
 \end{equation}
 An important difference between the canonical ensemble and the grand canonical ensemble corresponds to the possibility that in the latter can occur a HP first order phase transition to thermal AdS. Essentially, due to quantum effects a black hole can emits energy to external background through Hawking radiation and this allows to reach the thermal equilibrium between a stable black hole configuration and the thermal AdS space and its gran thermodynamic potential is zero. Fig. \ref{plotsG} summarizes the main results about this subject, and we can see there the HP phase transition between the stable black hole and thermal AdS, and also there is a second order phase transition between SMB and LBH at the cusp temperature. Moreover, by observing the slope and concavity of the thermodynamic potential, see Fig. \ref{plots2}, we can conclude that the black holes with $S < S_{cusp}$ (SBH) are thermally unstable and with $S > S_{cusp}$ (LBH), are thermally stable. Then, the former cannot establish the equilibrium with the thermal AdS space.

\newpage

\section{Concluding Comments}
\label{conclusion}

In this paper we studied the thermodynamics description of black hole solutions of Einstein theory coupled to a nonlinear power-law electromagnetic field. We considered the extended phase space, by considering the cosmological constant as a source of dynamical pressure, and we studied phase transitions of first order including the Hawking-Page phase transition, and second order phase transitions at both, canonical and grand canonical ensembles. Mainly, we found that first and second order phase transitions can occur at the canonical ensemble, while that for the grand canonical ensemble we found that phase transitions of first order, of Hawking-Page type can occur. However, the first order phase transitions depend on the parameter $k$, which describes the curvature of the spatial 2-section, yielding that in the canonical ensemble the first order phase transitions are allowed for $k=1$ (spherical), and in the grand canonical ensemble are allowed for $k=-1$ (hyperbolic). 
 
 For the canonical ensemble,  we obtained  that the temperature and the heat capacity are always positive with the only exception for the extreme black hole configuration. 
 From the analysis of the heat capacity at fixed charge we showed  that there is a region where  $C_{Q} < 0$, then according to Davies’s approach this indicates the presence of a type one phase transition, also it is important to remark that the temperature in this region is also negative and therefore it is considered as a non physical region, where the thermodynamic description breaks down. It is very well known that a black holes is a locally stable thermodynamic system if its heat capacity is positive and non vanishing. Besides, at the points where the heat capacity is  divergent there is a type two phase transition. Also, we have shown that the black hole solution is locally unstable from a thermal point of view, because the heat capacity $C_{Q}$ has a  divergent term, for the case where the spatial section is spherical ($k=1$). For the hyperbolic ($k=-1$) and flat ($k=0$) spatial sections  the  black hole configurations are  stable. Then, following Davies and according to Ehrenfest’s classification, second order phase transitions occur at those points where the heat capacity diverges. Then, we obtain three different phases, two stables phases where $C_Q>0$ for small and large horizon radius, and one unstable phase with $C_Q<0$ for intermediate radius. Therefore, there are three different phases SMB, MBH, and LBH.
  Also, we determined the Gibbs free energy, and we have shown that the critical pressure corresponds to
  the point where the system is undergoing a first order phase transition. For $P<P_C$ it appears the well known characteristic swallow tail where the first order phase transition  occurs at the intersection point between the SBH
  and LBH, with $T>0$ and  $C_Q>0$. Therefore, 
 SBH and LBM are locally stable or also we can interpreted as the phases with positive specific heat in the lower radius and higher radius regions are stable. When, $T>0$ and  $C_Q<0$ we have an unstable MBH configuration. This meta-stable configuration can be explained because states of the lowest Gibbs free energy are preferred for the system, similar situation can occur for some states on the small and large black hole branches.

From the grand canonical ensemble point of view, we found there is a critical point only for the case $k=-1$ and therefore there is a critical behavior similar to the  description in the canonical ensemble for $k=1$. For  cases $k=1$ and $k=0$,  the P-V analysis shows that there is no critical points because the condition $\partial P=\partial^2 P=0$ can not be satisfied. Finally, one important difference of the grand canonical ensemble in comparison to the canonical ensemble corresponds to the possibility to have a Hawking-Page first order phase transition to thermal AdS. Essentially, due to quantum effects a black hole can emit energy to the external background through Hawking radiation and this allows to reach the thermal equilibrium between a stable black hole configuration and the thermal AdS space and its grand thermodynamic potential is zero. 
Also, we have shown the existence of a HP phase transition between a stable black hole and the thermal AdS, and also there is a second order phase transition between SMB and LBH at the horizon radius where $C_{\Phi}$ is divergent, this point corresponds to the $T_{cusp}$ temperature.

  \acknowledgments
This work is supported by ANID Chile through FONDECYT Grant  No 1170279 (J. S.). Y.V. acknowledge support by the Direcci\'on de Investigaci\'on y Desarrollo de la Universidad de La Serena, Grant No. PR18142. B.W. was supported in part by NNSFC under grant No. 12075202.

\end{document}